# Valorization of biodigestor plant waste in electrodes for supercapacitors and microbial fuel cells

Bridget K. Mutuma[1], Ndeye F. Sylla[1], Amanda Bubu[1], Ndeye M. Ndiaye[1], Carlo Santoro[2], Alessandro Brilloni[3], Federico Poli[3], Ncholu Manyala[1]*, Francesca Soavi[3]**

[1] Department of Physics, University of Pretoria, Pretoria, South Africa

[2] Department of Materials Science, University of Milano-Bicocca, Via Cozzi 55, 20125 Milano, Italy

[3] Department of Chemistry "Giacomo Ciamician", Alma Mater Studiorum – Università di Bologna, Via Selmi 2, 40126, Bologna, Italy

Corresponding authors:

* **Ncholu Manyala:** ncholu.manyala@up.ac.za

** **Francesca Soavi:** francesca.soavi@unibo.it

Abstract

This study aims at demonstrating that wastes from anaerobic biodigester plants can be effectively valorized as functional materials to be implemented in technologies that enable efficient energy management and water treatment, therefore simultaneously addressing the Water-Energy-Waste Nexus challenges. Lignin, the main solid residue of the biodigester plant, has been valorized into activated biochar with a mild activation agent, like $KHCO_3$, to produce electrode of supercapacitors and microbial fuel cells. In addition, the same sludge that is the liquid effluent of the biodigester plant has been exploited as inoculum and electrolyte for the MFC. The lignin-derived carbons obtained at lignin/$KHCO_3$ mass ratios of 1:0.5 (LAC-0.5) and 1:2 (LAC-2) comprised of mesopores and micropores displaying BETs of 1558 $m^2g^{-1}$ and 1879 $m^2g^{-1}$, respectively. LAC-2 carbon exhibited a superior specific capacitance of 114 F $g^{-1}$ in 2.5 M $KNO_3$ with respect to LAC-0.5. A supercapacitor with LAC-2 electrodes was built displaying specific energy specific power up to 10 Wh $kg^{-1}$ and 6.9 kW $kg^{-1}$, respectively. Durability tests showed that the device was able to maintain a capacitance retention of 84.5% after 15,000 charge-discharge cycles. The lignin-derived carbons were also studied as electrocatalysts for ORR in a neutral medium. The LAC-2 showed higher ORR electrocatalytic activity than LAC-0.5. The interconnected porous network and the high surface area made the lignin-derived porous carbons suitable electrode materials for dual applications. The feasibility of the use of LAC 2 carbon incorporated in an air breathing cathode for MFC applications is also reported.

1. **Introduction**

Affordable and clean energy, clean water and sanitation, and responsible consumption are three of strategic goals of the 2030 Sustainable Agenda [1]. These goals are extremely interconnected within the so-called Energy-Water-Waste Nexus [2]. Indeed, energy is needed to provide clean water from wastewater and water is required in power plants. In turn, both energy and water are needed to treat wastes. Therefore, transforming wastes into energy-related valuable products closes the loop and is expected to be a powerful strategy to address sustainability agenda goals. The transformation and valorization of organic wastes into smart materials is envisioned to be implemented in technologies that enable an efficient management of renewable energy sources and an efficient treatment of waters, like electrical supercapacitors (even called electrical double layer capacitors, EDLCs) and microbial fuel cells (MFCs).

EDLCs store electric energy by an electrostatic process that gives rise to the so-called double layer capacitance at the interface between high surface area carbonaceous electrodes and the electrolyte. Their fast charge/discharge processes enable superior specific power and cycle life, but lower energy density, compared to batteries. For their fast response time, supercapacitors represent emerging technology for high peak power demanding applications, like electrical energy storage in renewable energy plants [3].

MFCs use electroactive microorganisms as biocatalysts to convert chemical energy stored in wastewater's organic matter into electricity. Bioanodes are coupled to air breathing cathodes, where electrocatalytic reduction of oxygen takes place. MFCs are alternative wastewater treatment devices. It was shown that MFCs are capable of degrading a plethora of simple organics and complex civil and industrial wastewaters [4]. However, MFCs deliver low power and are difficult to be directly used for practical applications. A solution is to couple MFCs with an energy storage system, like an EDLC, that accumulates the energy produced by the MFC and delivers it when needed at high power rate [5-8].

EDLCs and MFCs share high surface area carbon-based electrodes. Indeed, high surface area carbons are used in EDLCs to provide high specific capacitance (100-200 F g$^{-1}$) [3]. In turn, both MFC anode and cathode exploit carbonaceous materials in order to accommodate the electroactive biofilm and performing the oxygen reduction reaction (ORR), respectively [9].

Biomass is arising as a strategic and omnipresent carbon source. In recent years, the biochar obtained by pyrolysis and activation of biomasses has been widely explored for EDLC and MFC electrodes [10-11]. For EDLCs, the biochar porous architecture strongly affects electrode and device energy and power performance. High surface area has to be achieved by simultaneous tailoring pores size distribution for easy access of electrolyte ions, that is needed to set up the double layer capacitance [3]. A porous carbon matrix is created depending on the type and amount of activating agent, reaction time as well as the temperature [12-14]. For instance, high activation temperatures (> 600 °C), longer reaction time (>30 min) and the use of alkali hydroxides as activating agents is known to promote the creation of more pores on the carbon matrix [14-17]. $KHCO_3$ is a weak base that has been reported to result in porous carbons owing to its decomposition to $K_2CO_3$ at higher temperatures and it is capable to create pores resulting from the evolution of CO and $CO_2$ gases [16]. Compared to other activating agents, $KHCO_3$ is exceptional because, compared to KOH, it is less corrosive and more environmentally friendly Moreover, the decomposition of $KHCO_3$ produces more gas and creates a good expansion effect compared to KOH, which is beneficial to pore formation and production of hierarchically porous carbons [15-21].

One of the most challenging issues to be addressed for MFC application is the high cost of the electrodes material and the sluggish oxygen reduction reaction (ORR) occurring at the cathode [22]. Therefore, the development of low cost air breathing cathodes with low cost biochar material with high electrocatalytic activity is imperative. Typically, MFCs operate in a (circum)neutral medium and at room temperature to support the activity of the bacteria and their survival. At a neutral pH of 7, the low concentration of $H^+$ and $OH^-$ ($10^{-7}$ M) influences ORR kinetics and MFC power output [23-24]. Activated carbons provide large surface areas and hierarchical porous structure that allow for faster ORR kinetics and high-power output in microbial fuel cells [25].

Activated biochar has been obtained from a wide range of biomasses and lignin stands out because it is the third most abundant natural polymer. In addition, lignin is one of the major wastes of anaerobic digestion processes and pulp and paper making industries [26-28]. Its pyrolysis processes have been deeply investigated and lignin-derived biochar has already been proposed as an EDLC electrode component [11, 29-33]. Highly porous carbons have been derived from lignin using various chemical activating agents, including $KHCO_3$ [18]. However, the effect of lower mass ratios of lignin biomass/$KHCO_3$ on their physicochemical, electrochemical and electrocatalytic properties is rarely reported. Besides, the ability to use a lower and an appropriate mass ratio of lignin biomass/$KHCO_3$ is a plausible solution for environmental sustainability and the method is suitable for large scale production of activated carbons. In our previous studies, we exploited $KHCO_3$ to activated peanut shells derived biochar. We observed that the specific surface doubled by increasing the biomass to $KHCO_3$ mass ratio from 1:1 to 1:2. A further increase of the ratio only provided less than 10% gain in specific surface area [16].

This study emphasizes on the reusability of wastes from biorefinery industries and the use of smaller amounts of a less corrosive activating agent, like $KHCO_3$, as a suitable approach for sustainable cheap and easily processable porous carbons from lignin. The effect of $KHCO_3$ activating agent on the structural and textural properties of lignin-derived activated carbons is reported. The biochar obtained is then tested in EDLCs featuring a neutral aqueous solution of $KNO_3$ that offers the advantage of being greener with respect to organic electrolytes [3,16, 19]. In addition, the biochar is investigated as ORR electrocatalysts to be used in MFC cathodes. The same sludge that is the liquid effluent of the biodigester plant has been exploited as inoculum and electrolyte for the MFC.

## 2. Materials and Methods

### 2.1. Starting materials

Potassium hydrogen carbonate, $KHCO_3$ (99.99%), potassium nitrate, $KNO_3$ (99.99%), hydrochloric acid, HCl (37 %), carbon acetylene black (99.95%), polyvinylidene difluoride, PVDF (99 %) and N-methyl-2-pyrrolidone, NMP (99%), were purchased from Merck chemicals. Argon, Ar (99.99%) was used as received from Afrox. Polycrystalline nickel foam (surface area of 420

m$^2$g$^{-1}$ and 1.6 mm thickness, Alantum (Munich, Germany) and microfiber filter paper (0.18 mm thickness, ACE chemicals) were used for the electrode preparation.

## 2.2. Preparation of lignin-derived activated carbons

The dried lignin was washed with water/ethanol mixture and dried at 80°C for 12 h in an electric oven. Approximately 3 g of lignin were soaked for 24 h in a KHCO$_3$ solution comprising of 1.5 g of KHCO$_3$ in 40 mL of deionized water. The mixture was then dried at 80°C for 12 h in an electric oven. The dried mold was transferred into a horizontal tubular furnace and gradually heated to 850 °C at a ramp rate of 5 °C min$^{-1}$ for 1 h under 250 cm$^3$ min$^{-1}$ flow of Ar gas. The obtained product was soaked in 3 M HCl for 8 h, washed with deionized water until a neutral pH was achieved and thereafter dried overnight at 80°C. The dried product was labelled as LAC-0.5. A similar procedure was repeated for mass ratio of 1:2 (3 g lignin, 6 g of KHCO$_3$ and 40 mL of water) and product labelled as LAC-2. The morphology of the carbons was evaluated using scanning electron microscopy (SEM) and transmission electron microscopy (TEM). The structural and textural properties of the samples were investigated using Raman spectroscopy, powder X-ray diffraction (XRD) and nitrogen physisorption analysis (Brunner, Emmet and Teller; BET).

## 2.3 Electrode preparation for supercapacitors

The electrode materials were prepared by mixing the active material, carbon acetylene black and polyvinylidene fluoride at a ratio of 8:1:1 followed by the addition of N-methyl-2-pyrrolidone (NMP) solution to form a slurry. The slurry was coated onto a nickel foam (1 cm × 1 cm) and dried at 80 °C for 12 h. Three electrode measurements were carried out using a Bio-Logic VMP300 potentiostat (Knoxville TN 37, 930, USA) controlled by the EC-lab V 11.40 software. A glassy carbon counter electrode, Ag/AgCl reference electrode and the LACs as working electrodes were used to perform the electrochemical measurements in a 2.5 M KNO$_3$ electrolyte solution.

A symmetric device was fabricated in a coin-cell type configuration using a fiberglass filter separator (Whatmann GF/F, thickness 360 μm) and an active material mass loading of approximately 6 mg cm$^{-2}$ in 2.5 M KNO$_3$. The cyclic voltammetry (CV) and galvanostatic charge-discharge (GCD) measurements were investigated at different scan rates and specific current values, respectively. The electrochemical impedance spectroscopy (EIS) measurements were carried out in a frequency range of 10 mHz to 100 kHz at an open circuit potential. The specific capacitances for a half-cell and single electrode of the symmetric device were calculated from the reciprocal of the slope ($\Delta V/\Delta t$) of the discharge curve of the GCD plots collected by a 3- and 2-electrode setup, respectively, using Eqs. 1 and 2 [33, 34]:

$$C_S = \frac{I\Delta t}{\Delta V m_s} \tag{eq. 1}$$

$$C_{el} = \frac{4I\Delta t}{\Delta V m_{tot}} \tag{eq. 2}$$

where I is the current applied, $\Delta t$ is the discharge time and $\Delta V$ is the potential window. In equation 3, $m_s$ is the mass of the electrode in a three-electrode configuration (half-cell) and the $C_S$ is the specific capacitance for a half-cell. In equation 2, $m_{tot}$ is the total mass of the positive and negative electrode in a two-electrode configuration (full cell/device) and the $C_{el}$ is the specific capacitance of a single electrode in a full-cell.

The specific energy ($E_s$) is expressed in J g$^{-1}$ as evaluated using Eq. 3. To convert the unit J g$^{-1}$ to Wh kg$^{-1}$ the expression in Eq. 3 was divided by a factor of 3600 and further multiplied by 1000 as illustrated in Eq. 4:

$$E_d = 0.5 C_s (\Delta V) \quad \text{(J g}^{-1}\text{)} \tag{eq. 3}$$

$$E_d = 0.5 C_s (\Delta V) = \frac{1000 \times C_{el}(\Delta V)^2}{2 \times 4 \times 3600} \quad \text{(W h kg}^{-1}\text{)} \tag{eq. 4}$$

Thus, the specific energy and corresponding specific power of the device were calculated according to Eqs. 5 and 6, where $IR_{drop}$ is the internal voltage drop:

$$E_S = \frac{C_{el}(\Delta(V - IR_{drop}))^2}{28.8} \tag{eq. 5}$$

$$P_S = \frac{3600 E_S}{\Delta t} \tag{eq. 6}$$

The value expression in Eq. 4 was multiplied by 3600 to convert the discharge time from h to seconds. In Eq. 4, the value of $C_{el}$ that was measured at each current density was used.

## 2.4 Oxygen reduction reaction kinetics

Rotating disk electrode (RDE) technique was used to investigate the ORR kinetics at the carbon surfaces. The LAC-0.5 and LAC-2 inks were prepared by suspending 8 mg of each catalyst separately into 1 mL of Ethanol-Water-5%Nafion solution (67:30:3 volume ratio) and sonicated for 1 min for ensuring a good dispersion. The catalyst loading investigated during this study was 0.56 mg cm$^{-2}$. A neutral electrolyte solution comprising of 0.1 M potassium phosphate buffer and 0.1 M KCl was purged with pure oxygen for 30 min prior to measurements. Ag/AgCl and Pt were used as reference and counter electrodes, respectively. Electrode potential values are given vs. Normal Hydrogen Electrode (NHE, -0.198 V vs. Ag/AgCl). Linear sweep voltammetry (LSV) measurements were used to compare the activity of the different catalysts for the oxygen reduction reaction. LSV was run from 0.6 V to -0.9 V at a scan rate of 5 mV s$^{-1}$ with a rotation speed varying from 400 rpm to 2500 rpm.

Koutecky-Levich equation is often used for calculating the number of electrons transferred during the ORR by the catalyst of interest (n) and it is described in equation 7:

$$\frac{1}{j} = \frac{1}{j_k} + \frac{1}{j_d} = \frac{1}{0.62\, nFC_{O_2} D_{O_2}^{2/3} v^{-1/6} \omega^{-1/2}} = \frac{1}{j_k} + \frac{1}{B\omega^{1/2}} \qquad (eq.\,7)$$

where j is the measured current density, $j_k$ is the electrode potential dependent kinetic current density of the ORR, $j_d$ is the diffusion-limited current density, $n$ is the average number of electrons transferred per catalytic activity, F is the Faraday's constant (96,485 C mol$^{-1}$), $C_{O_2}$ is the concentration of $O_2$ in the electrolyte (1.117 10$^{-6}$ mol mL$^{-1}$), $D_{O_2}$ is the $O_2$ diffusion coefficient in aqueous media (1.9 10$^{-5}$ cm$^2$s$^{-1}$), $A$ is the electrode geometric area, $v$ is the kinematic viscosity (0.01073 cm$^2$s$^{-1}$) and ω is the electrode rotation speed. The values used in this work were retrieved by [23, 35]

## 2.5. Air breathing MFC cathode fabrication and testing

The feasibility of the use of lignin-derived biochar as MFC cathode was evaluated in real environment. Specifically, a cubical-shape single chamber membraneless MFC (3 cm$^3$) was assembled with symmetrical electrodes. The electrodes featured lignin biochar, conductive carbon pure black additive and PTFE based binder, in 8:1:1 mass ratio. The electrodes were prepared with the procedure described in our previous work and pressed over a Ti current collector grid with a pressure of 1 $ton\ cm^{-2}$ [8]. The cathode had a mass loading of 60 mg and an area of 1.13 cm$^2$ (53 mg cm$^{-2}$). A solution composed of 50% phosphate buffer saline solution (1x) and 50 % sludge in volume with 160 mg L$^{-1}$ of sodium acetate was used as inoculum and MFC fuel. The sludge was the main liquid effluent of the biodigester plant (Biotech sys. S.r.l.), from which lignin waste was recovered.

The ORR performance of the lignin derived carbon was evaluated by linear sweep voltammetry (LSV). The cathode was used as the working electrode, the anode was used as the counter electrode, the reference electrode was an Ag/AgCl electrode. Electrode potential values are given vs. NHE.

## 3. Results

## 3.1. Morphological analysis

The morphology of the LAC samples was confirmed using scanning electron microscopy (SEM) and transmission electron microscopy (TEM). Figure 1a-b shows that the LAC samples are porous at low and high magnifications. It can be seen that in LAC-2, the higher amount of KHCO$_3$ generates more pores that are better interconnected than in LAC-0.5. The generation of a porous network can be ascribed to the decomposition of KHCO$_3$ at higher temperatures releasing K$_2$CO$_3$, CO$_2$ and H$_2$O [17]. The K$_2$CO$_3$ can further react with the carbon atoms releasing CO$_2$ gas and potassium compounds which create additional pores on the carbon matrix [13]. Besides, the K$_2$CO$_3$ can decompose at higher temperatures (> 700 °C) to yield K$_2$O and CO$_2$ gas and form more pores on the surface. These potassium compounds can intercalate into the carbon matrix in the form of K$_2$O which can be readily removed by washing with concentrated HCl. The porous network of the lignin-derived activated carbons was confirmed by the TEM images (Figures 1c-d).

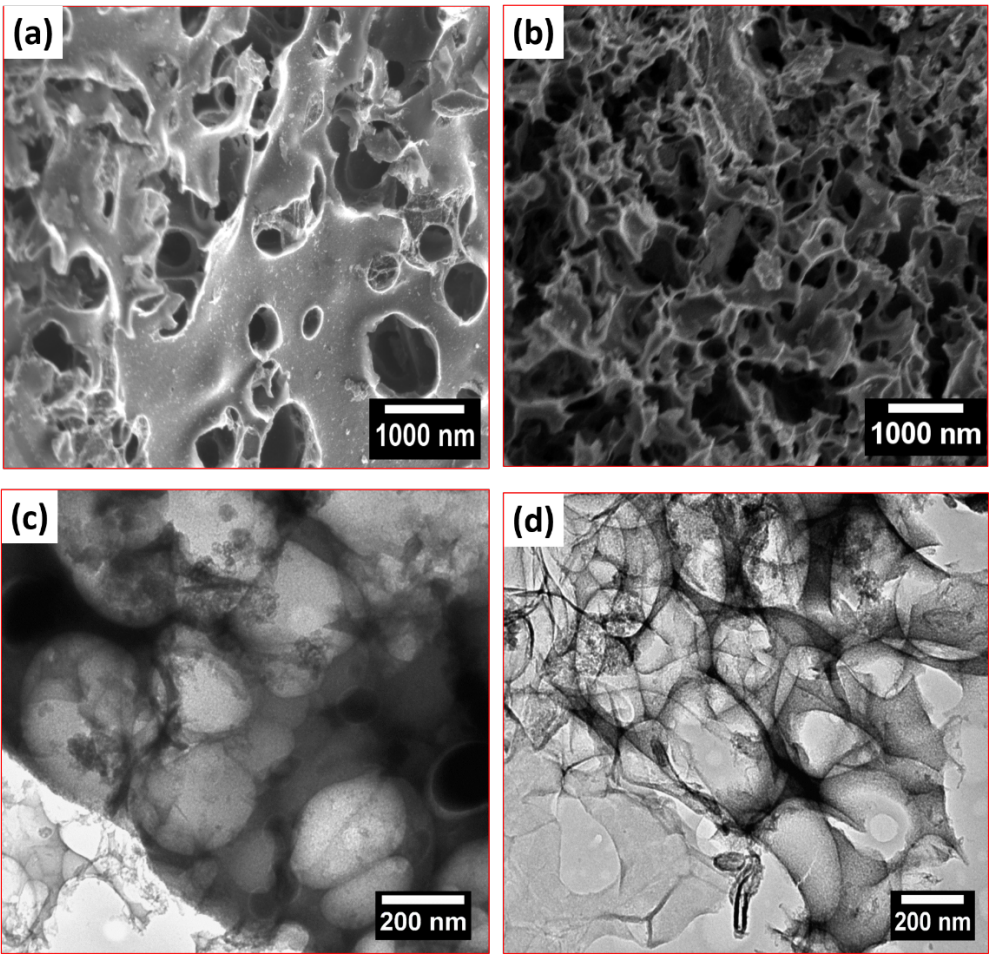

**Figure 1.** (a-b) Low magnification SEM images for (a) LAC-0.5 and (b) LAC-2; and (c-d) TEM images for (c) LAC-0.5 and (d) LAC-2, respectively.

## 3.2. Structural and textural properties

Raman spectroscopy was used for evaluating the degree of graphitization of activated carbons. Raman analysis for the LAC samples showed the presence of D peaks at 1331 – 1340 cm$^{-1}$ and G peaks at 1590 – 1594 cm$^{-1}$, respectively (Figure 2a). The broad D and G peaks indicated that the carbons were amorphous which is a typical characteristic of activated carbons. The D peak is ascribed to the breathing mode of sp$^2$ carbon due to the presence of sp$^3$ amorphous domains and defects within the carbon lattice, while the G peak is due to the bond stretching of sp$^2$ hybridized carbons [36-37]. The full-width half-maximum (FWHM) of the G peak was used to show amorphous carbon defects as a measure of bond distortion associated with the structural disorder while the FWHM of D peak was used to evaluate the basal plane defects on the materials as a measure of structural defects [38-39]. The FWHM of the D peak increased with increase in the amount of $KHCO_3$ indicating the creation of structural defects at higher $KHCO_3$ content. On the other hand, the FWHM of G peak varied for the two LAC samples indicating a slight difference in the amorphous carbon defects owing to bond angle and bond length distortions.

Interestingly, a D* peak was observed for the LAC-2 sample demonstrating the presence of $sp^2$-$sp^3$ carbon domains in the form of dangling bonds [40]. The $I_D/I_G$ of the activated carbons was calculated from the intensities of the D to G peaks and was found to be 0.94 ± 0.02, and 0.98 ± 0.02 for the LAC-0.5 and LAC-2, respectively (Table 1). This showed that all the carbons had a moderate degree of graphitization. Figure 2b shows the diffraction patterns of the LAC samples with broad peaks observed at 2θ = 23° and 44° corresponding to reflections of (002) and (100) planes of graphitic structures. The broad peaks suggest low crystallinity. Besides, it can be seen that with an increase in the amount of activating agent, the peaks broaden indicating the creation of a more disordered carbon matrix that is in agreement with the Raman data.

The nitrogen adsorption-desorption isotherms for all the samples are displayed in Figure 2c. All the samples exhibited type-IV isotherms with H3 hysteresis loops at 0.45 < P/P0 < 1.0, indicating the co-existence of micropores and mesopores. The specific surface areas were 1558 $m^2 g^{-1}$ and 1879 $m^2 g^{-1}$ for the LAC-0.5 and LAC-2 samples, respectively. The surface areas increased with increase in the amount of $KHCO_3$ revealing the formation of a more porous network at higher mass ratios. Figure 2d displays the pore size distribution of all the samples.

**Table 1.** Raman data and textural properties for the lignin-derived activated carbons

| Material | D peak position (cm$^{-1}$) | D peak FWHM (cm$^{-1}$) | G peak position (cm$^{-1}$) | G peak FWHM (cm$^{-1}$) | $I_D/I_G$ ratio (± 0.02) | Specific surface area (m$^2$g$^{-1}$) | Total pore volume (cm$^3$g$^{-1}$) |
|---|---|---|---|---|---|---|---|
| LAC-0.5 | 1340.0 | 167 | 1594.0 | 95 | 0.94 | 1558 | 0.46 |
| LAC-2 | 1335.9 | 185 | 1594.0 | 99 | 0.98 | 1879 | 0.75 |

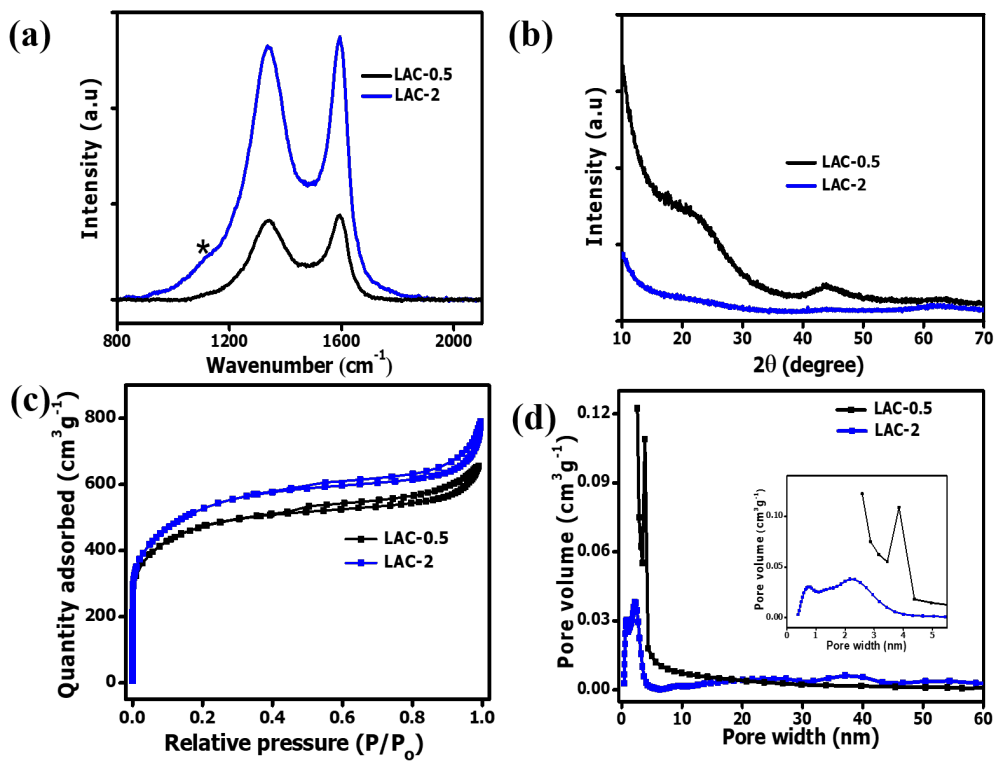

**Figure 2.** (a) Raman spectra, (b) XRD patterns, (c) N$_2$ adsorption-desorption isotherms and (d) pore size distribution plots for the activated lignin derived carbons.

At lignin:KHCO$_3$ mass ratio of 1:2, a large number of micropores and mesopores were generated compared to the ratio of 1:0.5 (inset) presumably because the intercalation of KHCO$_3$ in the carbon matrix took place to a larger extent. This is in agreement with the increase in total pore volume with the increasing amount of activating agent (Table 1).

### 3.3. Supercapacitors performance

Three electrode measurements were carried out for all the materials using a carbon counter electrode, Ag/AgCl reference electrode and 2.5 M KNO$_3$ electrolyte solution. The cyclic voltammograms of the LAC-0.5 and LAC-2 electrodes taken at varying scan rates featured a rectangular shape which is typical of an ideal EDLC behavior, both in the negative (-0.8 V to 0 V vs. Ref.) and positive (0 V to +0.8 V vs. Ref.) potential window (Figures 3a-b and 4a-b). Figures 3c-d and 4c-d show the galvanostatic charge-discharge plots in the negative and positive potential ranges for LAC-0.5 and LAC-2 electrode materials. The GCD plots at varying specific currents (1, 2, 3, 4, 5 and 10 A g$^{-1}$) displayed triangular shape, therefore confirming the capacitive behavior. LAC-2 exhibited a longer discharge time compared to LAC-0.5. This revealed better charge storage capability for the LAC-2 electrode material. The specific capacitance of the electrodes ($C_s$) was calculated from the GCD plots using Eq. 1.

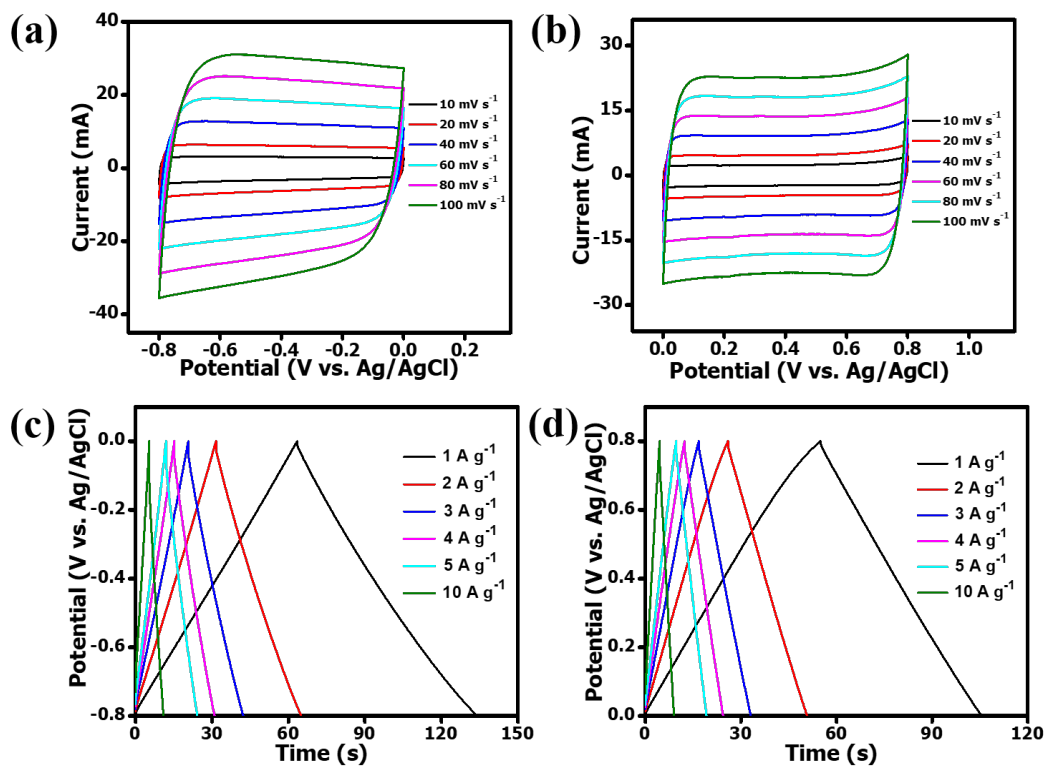

**Figure 3.** (a-b) CV curves in the potential ranges (a) -0.8 V to 0 V vs. Ag/AgCl and (b) 0 V to +0.8 V vs. Ag/AgCl at different scan rates, and (c and d) galvanostatic charge-discharge plots in both the negative and positive potential range at different specific currents of the LAC-0.5 electrode material in 2.5 M KNO$_3$ electrolyte.

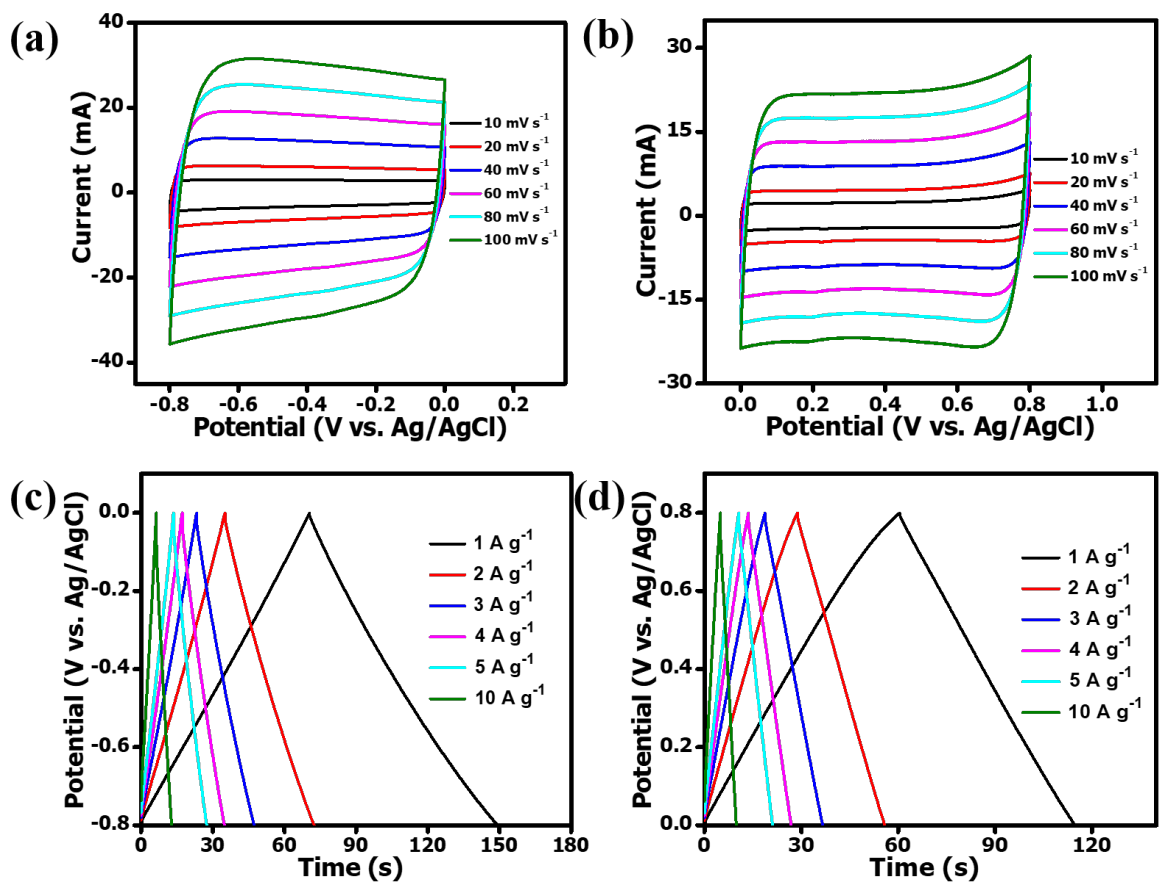

**Figure 4.** (a) CV curves in the potential ranges (a) -0.8 V to 0 V vs. Ag/AgCl and (b) 0 V to +0.8 V vs. Ag/AgCl at different scan rates, and (c and d) galvanostatic charge-discharge plots in both the negative and positive potential range at different specific currents of the LAC-2 electrode material in 2.5 M $KNO_3$ electrolyte.

Figure 5a and 5b display the $C_s$ values plotted as a function of specific current taken both in the negative and the positive potential range. All the electrodes exhibited good rate capabilities for the specific current values of 1 A $g^{-1}$ to 10 A $g^{-1}$. The specific capacitance values for the LAC-0.5 and LAC-2 at 1 A $g^{-1}$ were found to be 64 and 69 F $g^{-1}$, respectively, in the positive potential range (Figure 5a). In the negative potential window, the values were 89 and 100 F $g^{-1}$ for the LAC-0.5 and LAC-2, respectively (Figure 5b). For the various specific currents, LAC-2 electrode material exhibited higher specific capacitance values than the LAC-0.5. This can be ascribed to the higher specific surface area and pore volume as reported by the textural properties and the porous carbon network morphology of the LAC-2 material. In addition, the disordered carbon and structural defects could enhance the surface wettability of the LAC-2 electrode leading to higher specific capacitance values.

The capacitive behavior of the electrode materials was further investigated using electrochemical impedance spectroscopy (EIS). Figure 5c displays the Nyquist plots for the LAC sample electrodes with a quasi-vertical line parallel to the imaginary $Z''$-axis in the low frequency region indicating a capacitive nature of the samples. LAC-2 displayed a shorter diffusion length suggesting a faster ion diffusion at the electrode-electrolyte interface compared to LAC-0.5 (Figure 5c inset). Ideally, electrolyte ion transport highly depends on the electrolyte concentration, carbon electrode surface porosity and the pore diameter [3] . Thus, the higher surface area and pore volume of LAC-2 could provide an efficient ion-accessible surface area for faster ion diffusion. The real-axis intercept of the Nyquist plot at the highest frequency, was measured to be 1.25 Ω and 1.07 Ω for the LAC-0.5 and LAC-2, respectively. Given that EIS was collected by a 3-electrode setup, these values are affected by the cell geometry, namely by the distance of the working electrode from the reference. Moreover, it is affected by the ionic resistance of the bulk electrolyte and electronic resistance of the carbon electrode. The latter includes the carbon/current collector and interparticle contact resistances. Generally, a lower resistance at the highest frequencies depicts a better electronic conductivity of the electrodes. Hence, the lower resistance value exhibited by LAC-2 suggested a better electronic conductivity than LAC-0.5. Besides, from the Raman data LAC-2 portrayed a moderate degree of graphitization with the presence of D* peak that is ascribed to the presence of defects in the form of dangling bonds. An electrode material comprising a high amount of disordered carbons results in better aqueous electrolyte ion affinity and good surface hydrophilicity.

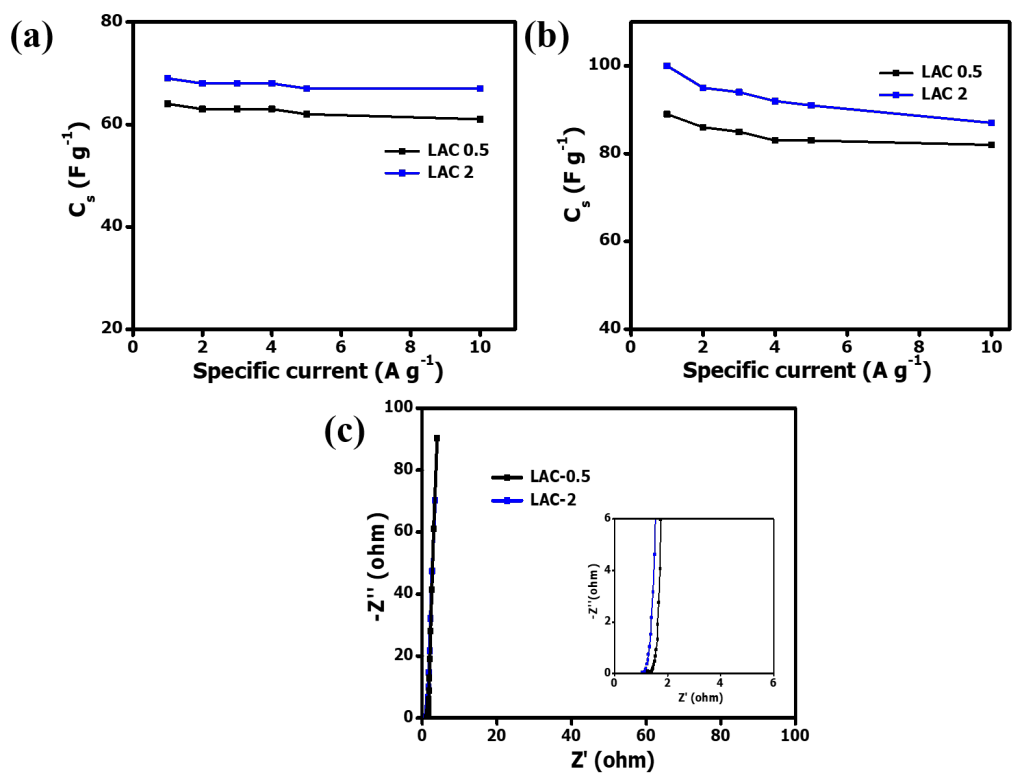

**Figure 5.** (a-b) Specific capacitance plotted as a function of specific current taken at an electrode potential range of (a) 0 V to +0.8 V vs. Ref, (b) -0.8 V to 0 V vs. Ref and (c) Nyquist plots of the LAC-0.5 and LAC-2 electrodes in 2.5 M KNO$_3$ electrolyte (3-electrode setup).

Based on the higher electrochemical performance of LAC-2, a symmetric device consisting of LAC-2 electrodes was fabricated with a 2.5 M KNO$_3$ electrolyte solution. Figures 6a and 6b present the CV plots of the EDLC collected by a 2-electrode setup, at different scan rates ranging from 10 mV s$^{-1}$ to 100 mV s$^{-1}$ within a cell voltage of 1.6 V. The device exhibited a rectangular CV shape even at higher scan rates (Figure 6b) demonstrating a good rate capability of the electrode material. The GCD plots at varying specific currents displayed a small $\Delta V_{ohm}$ drop showing a low ESR (Figure 6c). The specific capacitance of a single electrode (C$_{el}$) in the device was calculated from the GCD plots using Eq. 2 and plotted as a function of specific current as shown in Figure 6d. A maximum C$_{el}$ of 114 F g$^{-1}$ was obtained at 0.5 A g$^{-1}$ and a value of 89 F g$^{-1}$ was maintained at a specific current of 10 A g$^{-1}$ showing a good rate capability of the device (Figure 6d). This value is comparable to values in the literature on lignin-derived activated carbon electrodes [29-33]. For instance, Saha *et al.* [33] reported a specific capacitance of 91.7 F g$^{-1}$ at 2 mV s$^{-1}$ in 6 M KOH electrolyte for porous carbons produced by KOH activation of lignin. Similarly, Navarro and co-workers investigated the electrochemical performance of nanoporous carbon derived from KOH activated lignin and obtained a specific capacitance of 87 F g$^{-1}$ at 0.1 A g$^{-1}$ in organic electrolyte [32]. The slightly higher values reported in this study can be linked to the creation of mesopores and micropores by the activation of lignin with KHCO$_3$ yielding high pore volume and surface area of the LAC-2 sample.

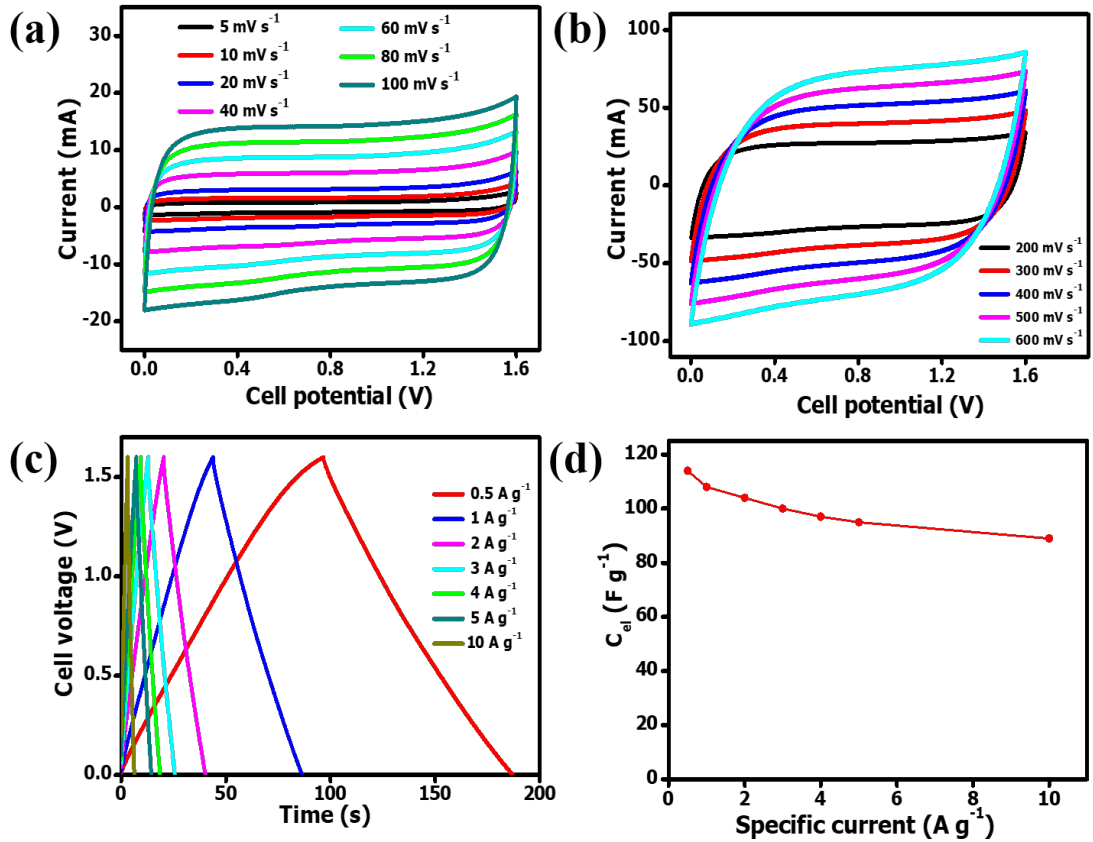

**Figure 6.** LAC-2//LAC-2 symmetric device: (a-b) CV plots at low scan rates and high scan rates, (c) charge-discharge curves and (d) specific capacitance as a function of specific current taken at a cell voltage from 0 V to 1.6 V (2-electrode setup).

Figure 7a presents a Ragone plot showing the specific power as a function of the specific energy of the device. The maximum specific energy was calculated using Eq. 3 and found to be 10 Wh kg$^{-1}$ with a corresponding specific power of 397 W kg$^{-1}$ at a specific current of 0.5 A g$^{-1}$. Besides, the specific power of 6.9 kW kg$^{-1}$ was achieved at a specific current of 10 A g$^{-1}$. To further evaluate the stability of the device, a cycling test was carried out on the device for up to 15,000 charge-discharge cycles. The device exhibited a coulombic efficiency of 99.84 % with capacitance retention of 84.5 % at a specific current of 5 A g$^{-1}$ (Figure 7b). This capacitance retention value was higher than the 80 % value reported by Jeon *et al.* [30] on lignin derived carbon after 10,000 cycles in 1 M $H_2SO_4$. The Nyquist plots of the device before and after cycling test were almost similar with the curve close to parallel with the imaginary Z″-axis indicating a capacitive behaviour (Figure 7c). The ESR was evaluated from the real-axis intercept at the highest frequencies and, after 15,000 cycle, it was slightly higher (1.73 Ω) than the one obtained before the cycling test (1.61 Ω) suggesting a possible increase in the carbon/current collector contact resistance after the cycling process. On the other hand, the high frequency semicircle diameters, that represent the interparticle electronic and ionic resistances, were almost similar: 0.11 Ω before cycling and 0.12 Ω after cycling. These results show that the device was quite stable even after 15,000 charge-discharge cycles.

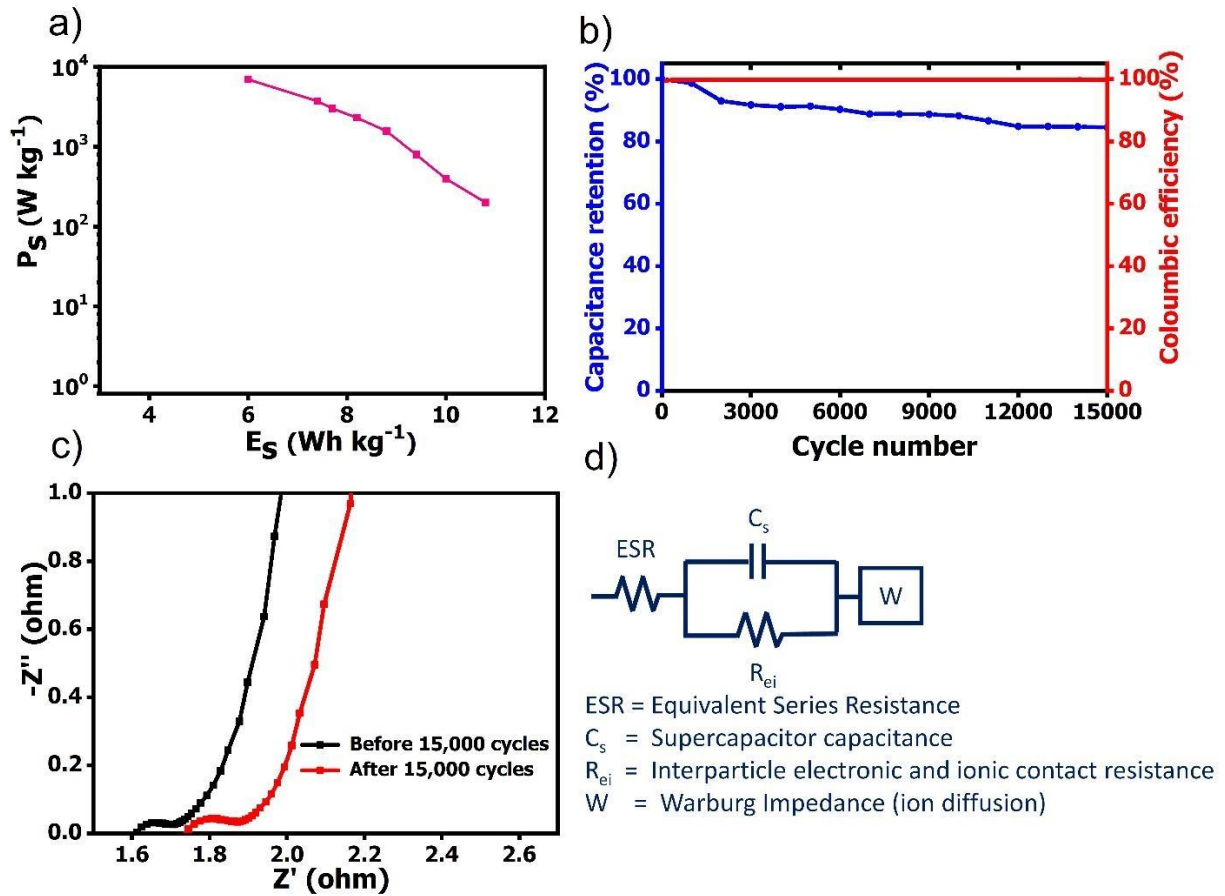

**Figure 7.** (a) Ragone plot, (b) cycling test showing capacitance retention and coulombic efficiency and (c) Nyquist plots before and after 15,000 charge-discharge cycles of the LAC-2 symmetric device (2-electrode setup) with the equivalent circuit that models the SC impedance.

### 3.4. Oxygen reduction reactions kinetics

The electrocatalytic activity towards ORR in neutral media of the two LAC materials obtained from pyrolyzed lignin was also tested. The polarization curves for the LAC catalysts obtained by RDE, at varying rotation speeds (400, 900, 1600 and 2500 rpm) are presented in Figures 8a-8b. The onset potential ($E_{on}$) for LAC-0.5 was 0.089 V (vs NHE) and for LAC-2 was slightly higher and corresponded to 0.122 V (vs NHE). For both LAC-0.5 and LAC-2 catalysts, no plateau corresponding to the mass transport limited current density was observed. This suggested that the ORR activity resulted from a mixed kinetic-diffusion controlled mechanism. An increase in the limiting current density with rotating speed was observed for all the catalysts indicating a higher oxygen flux from the bulk solution to the electrode surface at higher rotating speeds [41].

The current density values for the LAC-0.5 and LAC-2 were 6.23 and 8.53 mA cm$^{-2}$, respectively at -0.7 V vs NHE for a rotating speed of 2500 rpm. Along the entire potential window investigated from $E_{on}$ to -0.7 V vs NHE, LAC-2 outperformed LAC-0.5 having higher current densities at the same potentials. The differences in the ORR activity between the two catalysts can be linked to their morphologies, porosities and textural properties. For instance, LAC-2 material exhibited a well-interconnected porous network, a higher pore volume and a high surface area that can allow for exposure of the active sites to the reactants. On the other hand, the LAC-0.5 catalyst had lower pore volumes and less interconnectivity.

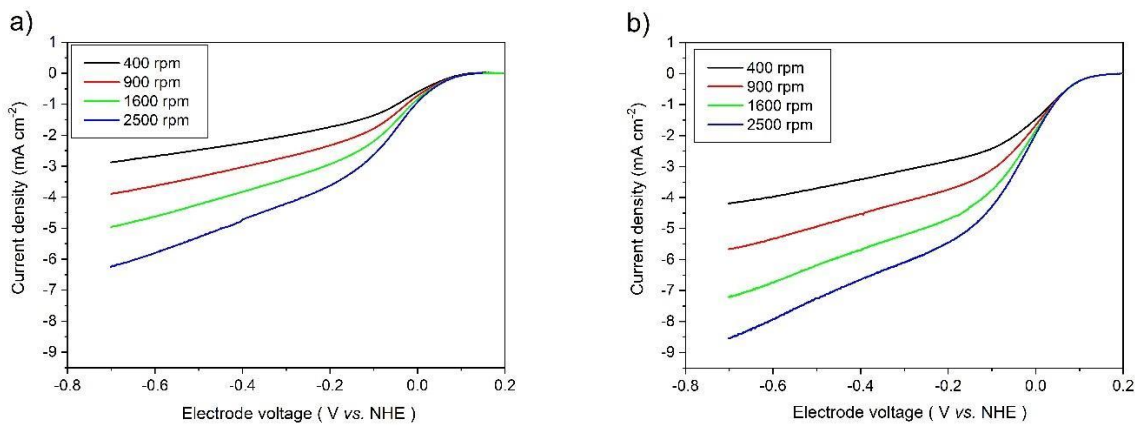

**Figure 8.** Rotating disk electrode plots for (a) LAC-0.5 and (b) LAC-2 with the catalyst loading of 0.56 mg cm$^{-2}$ at 5 mV s$^{-1}$ scan rate.

In order to calculate the number of electrons transferred during the ORR process, the Koutecky-Levich (K-L) graphs were plotted (Figure 9a and 9b). All the curves taken between 0 V and -0.2 V (vs. NHE) exhibited a linear and almost parallel trend suggesting that the activated carbons catalysts follow first-order kinetics [41]. From the intercepts of the linear curves in Figure 9a and 9b, the kinetic-limited current density ($j_k$) was extracted as well as the coefficient B (eq. 7). Knowing these two values, it was possible to calculate the number of electrons involved within ORR. Figure 9c reports the number of electrons transferred during ORR at different potentials, estimated by Koutecky−Levich (K-L) analysis.

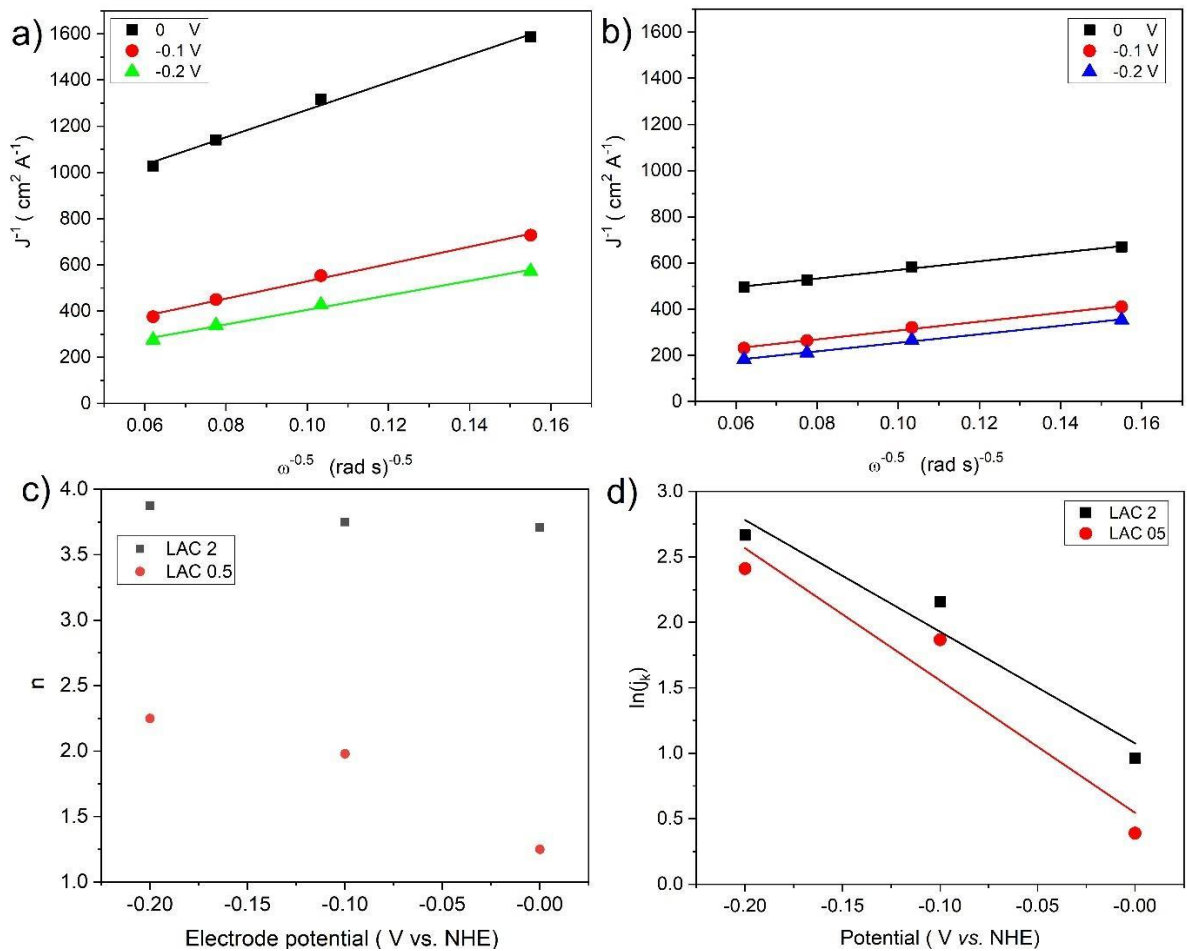

**Figure 9. (a-b)** K-L plots taken at different potential values of 0 V, -0.1 V and -0.2 V for LAC-0.5 (a) and LAC-2 (b), (c) number of electrons transferred estimated by K-L analysis and (d) Tafel plots.

It is well known that carbon-based catalysts are capable of reducing oxygen following a 2e- or a 2x2e- transfer mechanism [41]. In this specific case, LAC-0.5 seems to follow a straight 2e- transfer mechanism while LAC-2 instead seems to follow a mixed 2e- and 2x2e- transfer mechanism. In fact, the number of electrons transferred were much closer to 4 with values of 3.7, 3.76 and 3.92 for the potential of 0 V, -0.1 V and -0.2 V (vs NHE), respectively. This different pathway might be correlated to the porosity and the different pore size distribution of the two samples investigated. It is possible that larger cumulative porosity might be useful for providing more active sites for the ORR to occur. It must be underlined that the catalysts loading used for this study is quite high. In fact, it was shown that an increase in the catalyst loading led to higher performance and the peroxide intermediate is disproportionated within the thick catalyst layer itself [42-43]. It was shown that simple unmodified carbon black was able to perform a 2e-reduction of oxygen. On the contrary, the activated carbon had a higher electron transfer mechanism and superior performance [44]. In Watson et al. the authors screened nine different commercially available activated carbons finding values of n between 2 and 3.5 [45]. Particularly, it was shown that overall porosity and the pore size distribution was related to the number of electrons transferred and the overall electrocatalytic activity. In another recent study, Pepè Sciarria et al. used activated biochar as a cathode catalyst for ORR finding the number of electrons transferred in the range between 2.9 and 3.9 [46].

Tafel curves were calculated in the range 0 V to -0.2 V (vs NHE) and plotted in Figure 9d. The cathodic transfer coefficient βn was calculated from the Tafel slope using the following equation [47] below:

$$\ln \ln (j_k) + \frac{\beta nF}{RT}(E - E^0) \tag{8}$$

where $j_o$ is the exchange current density, R is the gas constant 8.314 J mol$^{-1}$ K$^{-1}$, and T is the standard room temperature. The cathodic transfer coefficient (**βn**) for LAC-0.5 and LAC-2 electrocatalysts was found to be 0.27 and 0.23 ($\pm$ 20% ), as evaluated within 0 V and -0.2 V vs NHE. At the lowest overpotentials, i.e. between 0 and -0.1 V vs. NHE, the transfer coefficients resulted in 0.31 and 0.36 for LAC-0.5 and LAC-2, respectively. For both catalysts, at a loading of 0.56 mg cm$^{-2}$, the cathodic transfer coefficient was similar to values reported on glassy carbon electrodes in the neutral medium [48]. For the LAC-0.5 catalyst, slightly higher cathodic transfer coefficients were recorded as compared to the LAC-2. The differences in the cathodic transfer coefficient for the LAC catalysts can be ascribed to the variations in the pore diameter, pore length and electronic conductivity of the samples. It can be concluded that increasing the amount of KHCO$_3$ resulted in porous carbons with a larger number of structural defects, more pores and consequently higher surface area and pore volumes which led to an increase in the active sites and regions where the ORR can occur with a positive impact on their electrochemical performance. Overall, LAC-2 sample exhibited superior electrocatalytic properties as well as electrochemical performance owing to its unique porous network, a high pore volume, high specific surface area as well as unique structural properties.

## 3.5. MFC air breathing cathode operating in neutral media

The ORR electrocatalytic activity of the LAC 2 carbon was further validated in real environment, i.e. in an air breathing MFC. The air breathing cathode was fabricated utilizing LAC 2 as an electrocatalyst. The fuel as well as inoculum were mainly based on the effluent of the anaerobic biodigester plant. Figure 10 reports the polarization curve obtained by linear sweep voltammogram of the LAC 2 MFC cathode performed at 0.2 mV s$^{-1}$. The Open circuit potential (OCP) is consistent with the values observed in Figure 8. The Figure compares the LSV of LAC-2 and other air-breathing cathodes composed by commercially available activated carbons (AC) previously presented [59-50]. These air breathing cathodes had catalyst loading of roughly 40 mg cm$^{-2}$, comparable with the one used in this study (53 mg cm$^{-2}$). Notably, the commercial ACs showed higher OCP and electrocatalytic activity at low current density compared to LAC-2. This might be due to the iron impurities previously identified in the commercial AC samples [51]. At higher current densities, LAC-2 demonstrated superior electrocatalytic activity compared to the commercial ACs. This is probably due to the much higher available surface area of LAC 2. These results indicate the feasibility of the use of lignin derived biochar for fabricating air breathing cathodes for MFC applications.

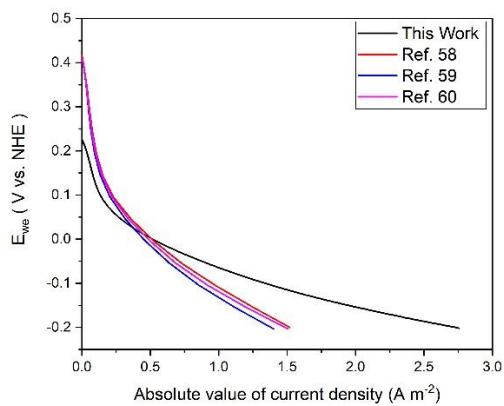

Figure 10. Polarization curve obtained by linear sweep voltammogram at 0.2 mV s$^{-1}$ of the LAC-2 MFC air breathing cathode (black curve) and commercial activated carbons: red line curve is imported from [48], Elsevier under license CC BY 4.0 (https://creativecommons.org/licenses/by/4.0/), the curve in blue line was imported from [49], Elsevier, under license CC BY 4.0 (https://creativecommons.org/licenses/by/4.0/), the green line

## 4. Conclusions

Porous carbons derived from lignin were obtained by activating with $KHCO_3$ at varying mass ratios of 1:0.5 and 1:2. A lignin biomass: $KHCO_3$ mass ratio of 1:2 was found to be the optimal ratio for high specific surface area (1879 $m^2g^{-1}$) and high pore volume (0.75 $cm^3g^{-1}$). The Raman data showed that LAC-2 had a higher amount of disordered carbon than the other LAC samples. The obtained materials were tested as electrodes for EDLC and as electrocatalysts for ORR in neutral media and integrated in air breathing cathode architecture for MFC applications. All the electrode materials displayed good reversible charge storage capability in both the negative (-0.8 V to 0 V vs. Ref) and the positive (0 V to + 0.8 V vs. Ref) potential windows in 2.5 M $KNO_3$ electrolyte. A symmetric EDLC fabricated using LAC-2 electrode material exhibited a good specific capacitance of 28.5 F $g^{-1}$, corresponding to an electrode specific capacitance of 114 F $g^{-1}$, with a specific energy of 10 Wh $kg^{-1}$ and a corresponding specific power of 397 W $kg^{-1}$ at a specific current of 0.5 $Ag^{-1}$ in a cell voltage of 1.6 V. Moreover, the device demonstrated good capacitance retention of 84.5 % after 15,000 cycles. As an electrocatalyst, LAC-2 showed higher current density values and superior ORR activity as compared to LAC-0.5. The number of electrons transferred during ORR was higher for LAC-2. Once integrated in an air-breathing cathode, the material exploited high electrochemical ORR activity especially at high current densities.

Overall, this study demonstrates that biochar derived from lignin waste of anaerobic biodigester plants, activated with a mild activation agent, like $KHCO_3$, features capacitive and ORR electrocatalytic properties that are comparable or even superior with those of commercially available carbons.

Therefore, our main result is the demonstration that waste can be effectively valorized as functional materials to be implemented in technologies that enable efficient energy management and water treatment, therefore simultaneously addressing the Water-Energy-Waste Nexus challenges. Indeed, the good supercapacitor performance and ORR electrocatalytic behavior of the lignin-derived carbons indicated a potential use as cathode catalysts and electrode materials for microbial fuel cells and supercapacitors.

**Author Contributions**

B.K.M, N.F.S, N.M.N, and A.B. material preparation, characterization of materials and determination of electrochemical capacitor properties. B.K.M, A.B and F.P., determination of oxygen reduction reactions. A.B., C.S, interpretation of oxygen reduction reactions. B.K.M manuscript preparation and editing, C.S, F.S and N.M. critical revision and supervision of all aspects of the research. F.S and N.M., funding acquisition. All authors have read and agreed to the published version of the manuscript.

**Acknowledgments**

This research was carried out under the Italy-South Africa joint Research Program (ISARP) 2018–2020 (Italian Ministers of Foreign Affairs and of the Environment, Grant No. PGR00764) . The South African Research Chairs Initiative of the Department of Science and Technology and the National Research Foundation of South Africa (Grants No. 61056 and No. 113132) is also acknowledged. C. S. would like to thank the support from the Italian Ministry of Education, Universities and Research (Ministero dell'Istruzione, dell'Universita`e della Ricerca –MIUR) through the "Rita Levi Montalcini 2018"fellowship (Grant number PGR18MAZLI). F.S., F.P, and A.B. acknowledge funding from the European Union's Horizon 2020 research and innovation program under grant agreement No. 963550 (HyFlow Project)..

**Declaration of competing interest**

The authors declare that they have no known competing financial interests that could have appeared to influence the work reported in this paper.

**References**

[1] https://sdgs.un.org/2030agenda
[2] https://www.unwater.org/water-facts/water-food-and-energy/

[3] F. Béguin, V. Presser, A. Balducci, E. Frackowiak, Carbons and Electrolytes for Advanced Supercapacitors, Adv. Mater. 26 (2014) 2219.

[4] P. Pandey, V.N. Shinde, R.L. Deopurkar, S.P. Kale, S.A. Patil, D. Pant, Recent advances in the use of different substrates in microbial fuel cells toward wastewater treatment and simultaneous energy recovery. Appl. Energy 168 (2016) 706.

[5] C. Santoro, F. Soavi, A. Serov, C. Arbizzani, P. Atanassov, Self-powered supercapacitive microbial fuel cell: the ultimate way of boosting and harvesting power. Biosens. Bioelectron. 78 (2016) 229.

[6] C. Santoro, X.A. Walter, F. Soavi, J. Greenman, I. Ieropoulos. Self-stratified and self-powered micro-supercapacitor integrated into a microbial fuel cell operating in human urine. Electrochim. Acta 307 (2019) 241-252

[7] F. Soavi, C. Santoro. Supercapacitive Operational Mode in Microbial Fuel Cell. Curr. Opin. Electrochem. 22 (2020) 1.

[8] F. Poli, J. Seri, C. Santoro, F. Soavi. Boosting microbial fuel cells performance by the combination of an external supercapacitor: an electrochemical study. ChemElectroChem 7 (2020) 893

[9] C. Santoro, C. Flores Cadengo, F. Soavi, M. Kodali, I. Merino-Jimenez, I. Gajda, J. Greenman, I. Ieropoulos, P. Atanassov. Ceramic Microbial Fuel Cells Stack: power generation in standard and supercapacitive mode. Sci Rep 8 (2018) 3281.

[11] Z. Bi, Q. Kong, Y. Cao, G. Sun, F. Su, X. Wei, X. Li, A. Ahmad, L. Xie, C. M. Chen. Biomass-derived porous carbon materials with different dimensions for supercapacitor electrodes: a review. J. Mater. Chem. A 7 (2019).

[12] D. Bergna, T. Hu, H. Prokkola, H. Romar, U. Lassi. Effect of Some Process Parameters on the Main Properties of Activated Carbon Produced from Peat in a Lab-Scale Process. Waste and Biomass. Valor. 11 (2019) 2837.

[13] C. Xia, SQ. Shi. Self-activation for activated carbon from biomass: theory and parameters. Green Chem. 18 (2016) 2063.

[14] W. Li, K. Yang, J. Peng, L. Zhang, S. Guo, H. Xia. Effects of carbonization temperatures on characteristics of porosity in coconut shell chars and activated carbons derived from carbonized coconut shell chars. Ind. Crops. Prod. 28 (2008) 190.

[15] J. Sahira, A. Mandira, P.B. Prasad, P.R. Ram. Effects of activating agents on the activated carbons prepared from lapsi seed stone. Res. J. Chem. Sci. 2231 (2013) 1.

[16] N. Sylla, N. Ndiaye, B. Ngom, D. Momodu, M. Madito, B. Mutuma, N. Manyala, Effect of porosity enhancing agents on the electrochemical performance of high-energy ultracapacitor electrodes derived from peanut shell waste. Sci. Rep. 9 (2019) 1.

[17] M. Sevilla, A.B. Fuertes. A green approach to high-performance supercapacitor electrodes: the chemical activation of hydrochar with potassium bicarbonate. ChemSusChem 9 (2016) 1880.

[18] Y. Xi, D. Yang, X. Qiu, H. Wang, J. Huang, Q. Li, Renewable lignin-based carbon with a remarkable electrochemical performance from potassium compound activation, Ind. Crop. Prod., 124 (2018) 747-754.

[19] F.O. Ochai-Ejeh, D.Y. Momodu, M.J. Madito, A.A. Khaleed, K.O. Oyedotun, S.C. Ray, N. Manyala. Nanostructured porous carbons with high rate cycling and floating performance for supercapacitor application. AIP Advances 8 (2018) 055208

[20] N.F. Sylla, B. Mutuma, A. Bello, T. Masikhwa, S. Lindberg, A. Matic, N. Manyala. Stable ionic-liquid-based symmetric supercapacitors from Capsicum seed-porous carbons. J. Electroanal. Chem. 838 (2019) 119-128

[21] F. Poli, D. Momodu, G. E. Spina, A. Terella, B. K. Mutuma, M. L., Focarete, N. Manyala, F. Soavi. Pullulan-ionic liquid-based supercapacitor: A novel, smart combination of components for an easy-to-dispose device. Electrochim. Acta, 338 (2020) 135872.

[22] H. Rismani-Yazdi, S.M. Carver, A.D. Christy, O.H. Tuovinen. Cathodic limitations in microbial fuel cells: an overview. J. Power Sources 180 (2008) 683.

[23] S. Rojas-Carbonell, K. Artyushkova, A. Serov, C. Santoro, I. Matanovic, P. Atanassov. Effect of pH on the activity of platinum group metal-free catalysts in oxygen reduction reaction. ACS Catal. 8 (2018) 3041.

[24] D. Malko, A. Kucernak, T. Lopes. In situ electrochemical quantification of active sites in Fe–N/C non-precious metal catalysts Nature Comm. 7 (2016) 1.

[25] H. Yuan, Y. Hou, I.M. Abu-Reesh, J. Chen, Z. He. Oxygen reduction reaction catalysts used in microbial fuel cells for energy- efficient wastewater treatment: a review. Mater. Horiz. 3 (2016) 382.

[26] J.H. Lora, W.G. Glasser, Recent industrial applications of lignin: a sustainable alternative to nonrenewable materials. J. Polym. Environ. 10 (2002) 39.

[27] P. Carrott, M.R. Carrott. Lignin–from natural adsorbent to activated carbon: a review. Bioresor Technol 98 (2007) 2301.

[28] H. Wang, Y. Pu, A. Ragauskas, B. Yang. From lignin to valuable products–strategies, challenges, and prospects. Bioresor. Technol. 271 (2019) 449.

[29] J.I. Hayashi, A. Kazehaya, K. Muroyama, A.P. Watkinson, Preparation of activated carbon from lignin by chemical activation, Carbon 38 (2000) 1873.

[30] J.W. Jeon, L. Zhang, J.L. Lutkenhaus, D.D. Laskar, J.P. Lemmon, D. Choi, M.I. Nandasiri, A. Hashmi, J. Xu, R.K. Motkuri. Controlling porosity in lignin-derived nanoporous carbon for supercapacitor applications. ChemSusChem 8 (2015) 428.

[31] D. Saha, Y. Li, Z. Bi, J. Chen, J.K. Keum, D.K. Hensley, H.A. Grappe, H.M. Meyer III, S. Dai, M.P. Paranthaman. Studies on supercapacitor electrode material from activated lignin-derived mesoporous carbon. Langmuir 30 (2014) 900.

[32] A.M. Navarro-Suárez, J. Carretero-González, V. Roddatis, E. Goikolea, J. Ségalini, E. Redondo, T. Rojo, R. Mysyk. Nanoporous carbons from natural lignin: study of structural–textural properties and application to organic-based supercapacitors. RSC Adv. 4 (2014) 48336.

[33] B. Moyo, D. Momodu, O. Fasakin, A. Bello, J. Dangbegnon, N. Manyala. Electrochemical analysis of nanoporous carbons derived from activation of polypyrrole for stable supercapacitors. J. Mater. Sci. 53 (2018) 5229.

[34] G. E. Spina, F. Poli, A. Brilloni, D. Marchese, F. Soavi. Natural Polymers for Green Supercapacitors. Energies 13 (2020) 3115.

[35] V.C.A. Ficca, C. Santoro, A. D'Epifanio, S. Licoccia, A. Serov, P. Atanassov. Effect of Active Site Poisoning on Iron− Nitrogen− Carbon Platinum-Group-Metal-Free Oxygen Reduction Reaction Catalysts Operating in Neutral Media: A Rotating Disk Electrode Study. ChemElectroChem 7 (2021) 3044-3055.

[36] A.C. Ferrari. Raman spectroscopy of graphene and graphite: disorder, electron–phonon coupling, doping and nonadiabatic effects. Solid State Commun. 143 (2007) 47. [38] M. Pimenta, G. Dresselhaus, M.S. Dresselhaus, L. Cancado, A. Jorio, R. Saito. Studying

disorder in graphite-based systems by Raman spectroscopy. Phys.Chem.Chem.Phys 9 (2007) 1276.

[37] A. Ferrari, J. Robertson. Resonant Raman spectroscopy of disordered, amorphous, and diamondlike carbon. Phys. Rev. B 64 (2001) 075414.

[38] A. Ferrari, S. Rodil, J. Robertson. Interpretation of infrared and Raman spectra of amorphous carbon nitrides. Phys. Rev. B 67 (2003) 155306.

[39] C. Casiraghi, A. Ferrari, J. Robertson. Raman spectroscopy of hydrogenated amorphous carbons, Phys Rev B 72 (2005) 085401.

[40] A. Sadezky, H. Muckenhuber, H. Grothe, R. Niessner, U. Pöschl. Raman microspectroscopy of soot and related carbonaceous materials: spectral analysis and structural information. Carbon 43 (2005) 1731.

[41] K. Kinoshita. Electrochemical Society, Electrochemical Oxygen Technology, Wiley, 1992

[42] C. Santoro, M. Kodali, S. Herrera, A. Serov, I. Ieropoulos, P. Atanassov. Power generation in microbial fuel cells using platinum group metal-free cathode catalyst: effect of the catalyst loading on performance and costs. J. Power Sources 378 (2018) 169.

[43] A. Bonakdarpour, M. Lefevre, R. Yang, F. Jaouen, T. Dahn, J.P. Dodelet, J.R. Dahn. Impact of Loading in RRDE Experiments on Fe–N–C Catalysts: Two-or Four Electron Oxygen Reduction? Electrochem Solid-State Lett. 11 (2008) B105

[44] I. Merino-Jimenez, C. Santoro, S. Rojas-Carbonell, J. Greenman, I. Ieropoulos, P. Atanassov. Carbon-based air-breathing cathodes for microbial fuel cells. Catalysts 6 (2016)

[45] V. Watson, C. Nieto Delgado, B.E. Logan. Influence of Chemical and Physical Properties of Activated Carbon Powders on Oxygen Reduction and Microbial Fuel Cell Performance. Environ. Sci. Technol. 47 (2013) 6704.

[46] T.P. Sciarria, M. Costa de Oliveira, B. Mecheri, A. D'Epifanio, J.L. Goldfarb, F. Adani. Metal-free activated biochar as an oxygen reduction reaction catalyst in single chamber microbial fuel cells. J. Power Sources 462 (2010) 228183.

[47] R. Taylor, A. Humffray. Electrochemical studies on glassy carbon electrodes: I. Electron transfer kinetics. J. Electroanal. Chem. Interf. Electrochem. 42 (1973) 347.

[48] C Santoro, M Kodali, S Kabir, F Soavi, A Serov, P Atanassov. Three-dimensional graphene nanosheets as cathode catalysts in standard and supercapacitive microbial fuel cell. J. Power Sources 356 (2017) 371-380

[49] C. Santoro, A. Serov, R. Gokhale, S. Rojas-Carbonell, L. Stariha, J. Gordon, K. Artyushkova, P. Atanassov. A family of Fe-NC oxygen reduction electrocatalysts for microbial fuel cell (MFC) application: relationships between surface chemistry and performances. Appl. Catal. B: Environ. 205 (2017) 24-33.

[50] C. Santoro, A. Serov, L. Stariha, M. Kodali, J. Gordon, S. Babanova, O. Bretschger, K. Artyushkova, P. Atanassov. Iron based catalysts from novel low-cost organic precursors for enhanced oxygen reduction reaction in neutral media microbial fuel cells. Energy Environ. Sci. 9 (2016) 2346-2353.

[51] M. Kodali, C. Santoro, S. Herrera, A. Serov, P. Atanassov. Bimetallic platinum group metal-free catalysts for high power generating microbial fuel cells. J. Power Sources 366 (2017) 18-26.